\documentclass[%
 aip,
 amsmath,amssymb,
 reprint,%
]{revtex4-1}

\usepackage{graphicx}% Include figure files
\usepackage{dcolumn}% Align table columns on decimal point
\usepackage{bm}% bold math
\usepackage{tabularx}
\usepackage{algpseudocode}
\usepackage{algorithm}

\usepackage[utf8]{inputenc}
\usepackage[T1]{fontenc}
\usepackage{mathptmx}
\usepackage{etoolbox}
\usepackage{xcolor}

\makeatletter
\def\@email#1#2{%
 \endgroup
 \patchcmd{\titleblock@produce}
  {\frontmatter@RRAPformat}
  {\frontmatter@RRAPformat{\produce@RRAP{*#1\href{mailto:#2}{#2}}}\frontmatter@RRAPformat}
  {}{}
}%
\makeatother
\begin{document}

\preprint{AIP/123-QED}

\title{Scattering-Based Structural Inversion of Soft Materials via Kolmogorov-Arnold Networks}
% Force line breaks with \\
\author{Chi-Huan Tung}
\author{Lijie Ding}
\affiliation{Neutron Scattering Division, Oak Ridge National Laboratory, Oak Ridge, 37831, Tennessee, United States}

\author{Ming-Ching Chang}
\affiliation{Department of Computer Science, University at Albany - State University of New York, Albany, 12222, New York, United States}    

\author{Guan-Rong Huang}
\affiliation{Department of Engineering and System Science, National Tsing Hua University, Hsinchu 30013, Taiwan}

\author{Lionel Porcar}
\affiliation{Institut Laue-Langevin, B.P. 156, F-38042 Grenoble Cedex 9, France}

\author{Yangyang Wang}
\author{Jan-Michael Y. Carrillo}
\author{Bobby G. Sumpter}
\email{sumpterbg@ornl.gov}
\affiliation{Center for Nanophase Materials Sciences, Oak Ridge National Laboratory, Oak Ridge, 37831, Tennessee, United States}

\author{Yuya Shinohara}
\affiliation{Materials Science and Technology Division, Oak Ridge National Laboratory, Oak Ridge, 37831, Tennessee, United States}

\author{Changwoo Do}
\author{Wei-Ren Chen}
\email{chenw@ornl.gov}
\affiliation{Neutron Scattering Division, Oak Ridge National Laboratory, Oak Ridge, 37831, Tennessee, United States}

\date{\today}% It is always \today, today,

\begin{abstract}
Small-angle scattering (SAS) techniques are indispensable tools for probing the structure of soft materials. However, traditional analytical models often face limitations in structural inversion for complex systems, primarily due to the absence of closed-form expressions of scattering functions. To address these challenges, we present a machine learning framework based on the Kolmogorov-Arnold Network (KAN) for directly extracting real-space structural information from scattering spectra in reciprocal space. This model-independent, data-driven approach provides a versatile solution for analyzing intricate configurations in soft matter. By applying the KAN to lyotropic lamellar phases and colloidal suspensions---two representative soft matter systems---we demonstrate its ability to accurately and efficiently resolve structural collectivity and complexity. Our findings highlight the transformative potential of machine learning in enhancing the quantitative analysis of soft materials, paving the way for robust structural inversion across diverse systems.
\end{abstract}

\maketitle

% \begin{quotation}
% The ``lead paragraph'' is encapsulated with the \LaTeX\ 
% \verb+quotation+ environment and is formatted as a single paragraph before the first section heading. 
% (The \verb+quotation+ environment reverts to its usual meaning after the first sectioning command.) 
% Note that numbered references are allowed in the lead paragraph.
% %
% The lead paragraph will only be found in an article being prepared for the journal \textit{Chaos}.
% \end{quotation}

\section{Introduction}
\label{sec:1}
Small-angle scattering (SAS) techniques, such as small-angle neutron scattering (SANS) and small-angle X-ray scattering (SAXS), are indispensable tools for characterizing the structures of soft materials \cite{ILL}. These techniques provide critical insights into the structural arrangement of complex systems, including polymers, surfactants, and colloidal suspensions, by measuring scattered intensities as a function of the wave vector $Q$. The inverse problem in SAS involves extracting real-space structural information from these scattering patterns in reciprocal space. To achieve this, the scattering function must be modeled as parameters that represent the real-space structure of the system, typically through analytical approximations \cite{ILL_16}, to facilitate regression analysis.

However, many important systems exhibit structures that cannot be captured accurately by traditional analytical models. In these cases, the complexity of the system's arrangement and the limitations of existing models create significant challenges for extracting real-space information. To overcome these challenges, we have developed various machine learning algorithms as mathematical frameworks for parameter inversion, bypassing the need for explicit analytical modeling \cite{GPR1, GPR2, dingGPR1, dingGPR2, dingGPR3, VAE, CNN1, CNN2}. These machine learning approaches have shown great potential in precisely obtaining physical information from controlled test data. However, when applied to real experimental data, they often encounter practical issues. 
% For instance, the scattering functions generated by machine learning models are typically in the format of vectors corresponding to evenly spaced and fixed sampling points for $Q$. 
For instance, the convolutional neural network (CNN) used discrete transposed-convolution operations to produce the \(Q\)-dependent features in the scattering spectrum, which necessitate a regular grid of evenly spaced \(Q\)-values.
In contrast, real experimental data often exhibit unevenly spaced $Q$ points due to variations in instrument configurations, collimation conditions, and other experimental factors.

To address these challenges, we propose a feature generation framework based on the Kolmogorov-Arnold Network (KAN) \cite{KAN1}. The framework leverages the Kolmogorov-Arnold representation theorem \cite{KAN2, KAN3}, which guarantees the existence of a continuous, smooth representation of multivariate functions. In the context of this theorem, the KAN model generates a continuous and differentiable scattering function that closely approximates scattering functions. This approach overcomes the limitations imposed by fixed-point output in conventional neural network models that relying on convolution operations, making it a powerful tool for SAS analysis, particularly when dealing with experimental data characterized by irregular and unevenly distributed $Q$-points. 

In this paper, we demonstrate the efficacy of the KAN-based function optimized for least square regression routine through two case studies. First, we apply the framework to the analysis of defective lamellar phases, illustrating how it can resolve structural distortions and quantify topological defect distributions. Second, we explore charged colloidal suspensions, where KAN successfully maps structure factors to effective interaction potentials, showing excellent agreement with both experimental data and molecular dynamics simulation results.

This work establishes KAN as a transformative tool for structural inversion in soft materials, providing a powerful alternative to traditional analytical models, and demonstrates its application in analyzing experimental data through curve fitting. The ability of the KAN to handle complex, real-world experimental data positions it as a crucial tool for advancing the quantitative spectral analysis of soft matter systems.

The remainder of this paper is organized as follows. In Section II, we present the methods used to develop the KAN-based generative function for scattering intensity. In Section III, we describe the application of the framework to the two case studies: defective lamellar phases and charged colloidal suspensions. Finally, in Section IV, we discuss the results and conclude with a summary of the work and potential future directions.

\section{Methods}
\label{sec:2}

\subsection{Kolmogorov-Arnold Network (KAN)}
%--------------------------------------------------
\begin{figure}[htbp]
\centerline{
  \includegraphics[width=\linewidth]{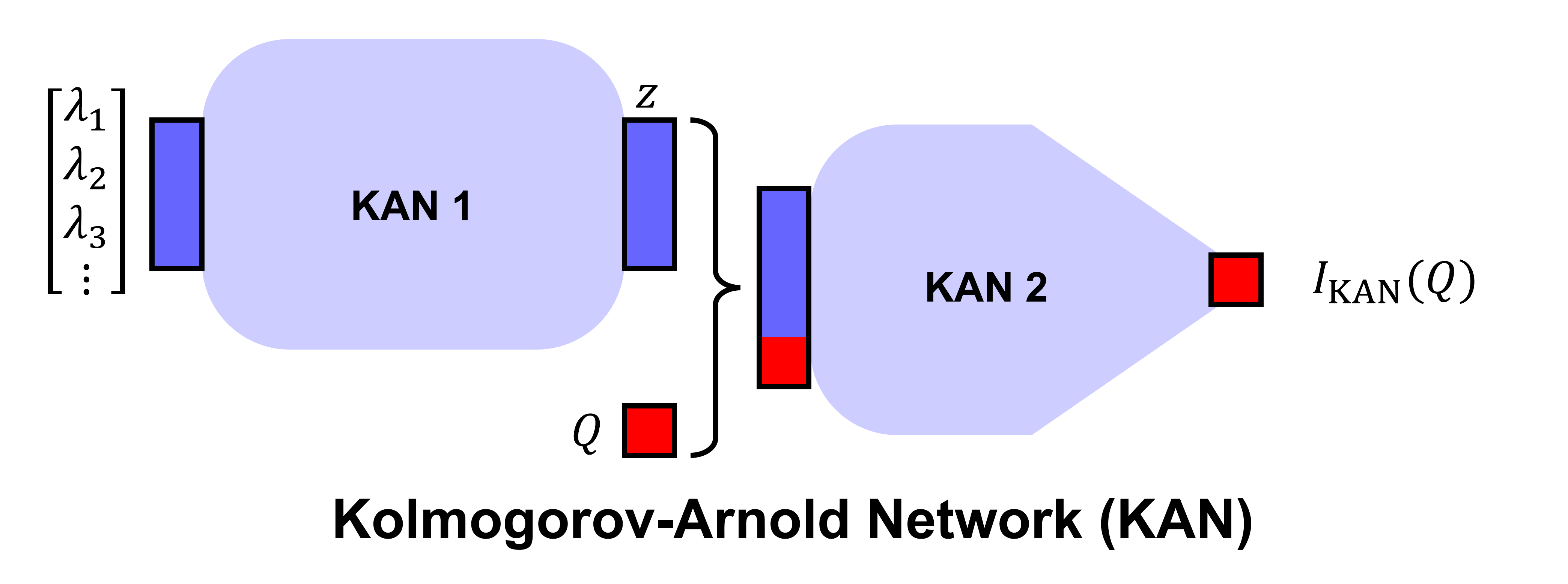}
  %\vspace{0 cm}
}  
\caption{A general setup of the KAN model setup for scattering functions. The model starts with input parameters \([\lambda_1, \lambda_2, \lambda_3]\), which pass through KAN layers (KAN 1) to generate latent variables \(z\). These are then combined with \(Q\) and processed by another set of KAN layers (KAN 2) to produce the final scattering intensity \(I_{KAN}(Q)\). }
\label{fig:2}
  %\vspace{-0.5cm}
\end{figure}
%----------------------------------------------------

% To address the challenges mentioned above, 
We choose to use KAN \cite{KAN1}, a versatile generative model that produces a continuous function for coherent scattering intensity \(I(Q)\). This approach overcomes the limitations of fixed-output CNN models, offering greater flexibility and better adaptability to real-world scattering data. Grounded in the Kolmogorov-Arnold representation theorem \cite{KAN2, KAN3}, KAN expresses any multivariate continuous function as a finite sum of univariate functions of input variables. In this framework, the multivariate \(Q\)-dependent scattering function is represented directly by simpler univariate components, avoiding the need for discrete convolution operations. Consequently, KAN eliminates the requirement for a fixed output grid, making it particularly well-suited for evaluating scattering intensities \(I(Q)\) at arbitrary \(Q\)-points, aligning seamlessly with experimental data.

The Kolmogorov-Arnold theorem \cite{KAN3} posits that any multivariate continuous function can be decomposed into a superposition of simpler univariate continuous functions. This decomposition facilitates the simplification of complex functions into more manageable components:
%--------------------------------------------------
\begin{equation}
f(x_1, \dots, x_n) = \sum_{q = 1}^{2n + 1} \chi_q \left( \sum_{p = 1}^n \psi^{pq} (x_p) \right),
\label{eq:1}
\end{equation}
%---------------------------------------------------
where \( f(x_1, \dots, x_n) \) represents the multivariate continuous function, and \( \chi_q \) and \( \psi^{pq} \) are univariate functions. Eqn.~\ref{eq:1} serves as a minimal example of two-layer KAN takes \(n\) inputs and produce one output. This nested structure can be extended to arbitrary depths and widths by adjusting the number of layers and the neurons within each layer.

The KAN-based generative model for SAS is depicted in Fig.~\ref{fig:2}. This model offers a more robust and flexible approach for regression analysis, especially when dealing with experimental scattering data. Similar to CNNs, the model begins with input parameters \([\lambda_1, \lambda_2, \lambda_3, \dots]\), which pass through an initial set of KAN layers (KAN 1) to generate latent variable set \(z\). To ensure that the output \(I(Q)\) explicitly depends on \(Q\), these latent variables are then combined with \(Q\) and processed through a second set of KAN layers (KAN 2) to produce the final output. Both KAN layers, as shown in Fig.~\ref{fig:2}, consist of nested, trainable functions parameterized by splines \cite{Spline}. The training process of spline coefficients is achieved by optimizing the spline coefficients to minimize the errors of KAN output using the batch optimization algorithm via backpropagation as described in our previous work~\cite{CNN2}.

Unlike CNNs, which are constrained by discrete transposed-convolution operations that producing discrete grid-based operations, KAN guarantees expressiveness through the Kolmogorov-Arnold representation theorem. Additionally, it avoids the checkerboard artifacts often caused by compounded transposed convolution layers~\cite{Checkerboard}. On the other hand, the smoothness of KAN with respect to \(Q\) is controlled through the selection of spline bases, which utilize piecewise low-degree polynomials with continuous derivatives. This approach ensures that the output values within each segment remain within a controlled range, effectively minimizing the potential for large oscillations. As a result, KAN generates the scattering intensity \(I(Q)\) as a smooth and continuous function over any chosen \(Q\)-points, without requiring interpolation. The flexibility of the KAN model can be fine-tuned by adjusting its depth, width, and the number of spline knots, depending on the complexity of the relationship between the parameters, \(Q\), and \(I(Q)\).

Note that Fig.~\ref{fig:2} only provides a general example of the KAN model setup for scattering functions. The detailed network architecture and the preparation of training sets for different soft matter systems in each specific scattering case study will be presented in the next section.

% In summary, the feasibility of CNN model is verified in controlled, evenly spaced test datasets, demonstrating the capability of solving the inverse problem of scattering with high numerical accuracy and computational efficiency. However, to match the general experimental data, the interpolation on CNN outputs is necessary, which can compromise data integrity, especially for experimental datasets with limited \(Q\)-resolution. In contrast, the KAN model treats scattering intensity as a continuous function, making it particularly numerically stable when handling experimental data with variable \(Q\)-spacing. By avoiding interpolation, KAN preserves data integrity and enhances regression accuracy, offering superior reliability for experimental datasets.

\subsection{Least Square Curve Fitting}

With KAN's capability to generate scattering functions based on given parameters, we can construct a least-squares curve-fitting algorithm to extract physical information from scattering experiments.

Algorithm~\ref{algorithm_curvefit} outlines a general iterative approach to gradually update the fitting parameters and the KAN-generated scattering function to align with the experimentally measured \(I_\mathrm{exp}(Q)\). At each step, the regression loss \(L_\mathrm{fit}\) is evaluated, with the goal of minimizing it until a predefined convergence condition is met. The core mechanism driving the parameter updates in this algorithm is a minimization algorithm. There are a variety of algorithms that can accomplish the task of parameter optimization, some of which, such as Quasi-Newton methods, rely on the gradient of the target function. This reliance poses limitations, particularly for CNNs, as their use as the foundation of gradient-based optimization algorithms in data analysis is hindered by the complexity introduced by activation functions between CNN layers. These activation functions make it challenging to compute the numerical derivatives of \(I(Q)\) with respect to the input parameters, reducing the feasibility of using CNNs in such applications. In contrast, since the KAN model only consists of additive operations on univariate continuous functions represented as splines, it allows for more efficient gradient evaluation. This capability unlocks the potential to utilize various numerical minimization methods, making KAN a more versatile choice for optimization tasks.

\begin{algorithm}[H]
\begin{algorithmic}[1]
\Require $I_\mathrm{exp}(Q)$ (experimental intensity), $f_{\mathrm{KAN}}(Q, \lambda)$ (KAN model function), $\lambda_{0}$ (initial guesses), $\mathrm{MaxIter}$ (maximum iterations), $L_c$ (convergence threshold)
\vspace{2mm}
\State Initialize $\lambda= \{ \lambda_i ; i=1\cdots N\} \gets \lambda_{0}$
\State Set iteration counter $i \gets 0$
\While{$i < \mathrm{MaxIter}$}
    \State Compute $I_{\mathrm{KAN}}(Q) \gets f_{\mathrm{KAN}}(Q, \lambda)$
    \State Compute the loss function: 
    \[
    L_{\mathrm{fit}} \gets \sum_Q \left\lvert I_\mathrm{exp}(Q) - I_{\mathrm{KAN}}(Q) \right\rvert^2
    \]
    \State Update $\lambda$ using a minimization algorithm to reduce $L_{\mathrm{fit}}$
    \State Increment iteration counter: $i \gets i + 1$
    \If{$L_{\mathrm{fit}} < L_c$}
        \State Set $\lambda_{\mathrm{ML}} \gets \lambda$
        \State \textbf{break}
    \EndIf
\EndWhile
\State \Return $\lambda_{\mathrm{ML}}$
\end{algorithmic}

\caption{\textbf{Least Squares Curve Fitting.}\\ This algorithm performs parameter extraction by fitting the experimental intensity data \( I_\mathrm{exp}(Q) \) with a model \( f_\mathrm{KAN}(Q, \lambda) \) generated curve using the least squares method. The algorithm iteratively updates them to minimize the loss function \( L_\mathrm{fit} \). A minimization algorithm is used to update \( \lambda \) at each iteration. Upon convergence, the optimized parameters \( \lambda_\mathrm{ML} \) are returned as the result.}
\label{algorithm_curvefit}
\end{algorithm}

\section{Results and Discussion}
\label{sec:3}
To demonstrate the feasibility of applying KAN to the inverse problem of soft matter structure characterization via SAS, we have deployed the least square curve fitting algorithm, Algorithm~\ref{algorithm_curvefit}\textbf{}, based on KAN architectures designed for two representative classes of soft materials: lamellar phases and colloidal suspensions. In the following, a detailed description of the KAN architecture is provided.

\subsection{Defective Lamellar Phases}

Lamellar phases are commonly encountered in numerous soft materials and biological systems such as block copolymers, surfactants, and liquid crystals. SAS techniques have been extensively used to characterize lamellar structures. Analytical models of scattering functions \cite{Roux, Nagle, Mortensen, Laggner, Ligoure, Monkenbusch, Porte3, Laggner2} assume perfect lamellar ordering, effective when layers remain uninterrupted. However, topological defects, often revealed by techniques like nuclear magnetic resonance \cite{Callaghan1, Davis, Chidichimo1, Chidichimo2, Photinos4, Coppola1, Halle0, Coppola2, Callaghan2, Topgaard1}, transmission electron microscopy \cite{Kleman1, Meyer, Kleman3, Costello1, Allain1, Allain2, Allain3, Kleman4, Strey1, Kleman5, Roux8, Jakli}, and conductivity measurements \cite{Photinos1, Boden0, Photinos2, Photinos3, Boden1}, frequently distort the layers, challenging the ideal model.

In analytical models, such as those assuming a square-wave density profile, the presence of perforations or crumpled surfaces in distorted lamellar phases is not explicitly accounted for. These phases can exhibit features similar to sponge-like structures, suggesting the need for a model that bridges the gap between ideal lamellar and sponge phases, with such transitions being controlled by the wave vector distribution. We propose a conformational descriptor for defective lamellar phases based on a wave field representation~\cite{CNN2}. In this framework, we have demonstrated that the structure of defective lamellar phases can be uniquely characterized by three parameters: $\sigma_k$, which represents the standard deviation of the wave vector in the wave field that models the density fluctuation of lamellar phases; $\Gamma$, which quantifies the degree of symmetry in the ordering of lamellar phases; and $\alpha$, which measures the volume ratio of amphiphilic molecules to water.

Within this descriptive context, solving the inverse scattering problem becomes more complex in terms of analytical modeling due to the highly non-linear relationship between the scattering function and these three parameters. To overcome these difficulties, we propose a machine learning approach based on Fig.~\ref{fig:2} to establish a generative function for scattering intensity as a function of these three parameters. A comprehensive library of two-point correlation functions for lamellar phases is constructed to train the generative KAN, thereby facilitating the inversion of real-space conformations from experimental scattering data.

%--------------------------------------------------
\begin{figure}[htbp]
\centerline{
  \includegraphics[width=\linewidth]{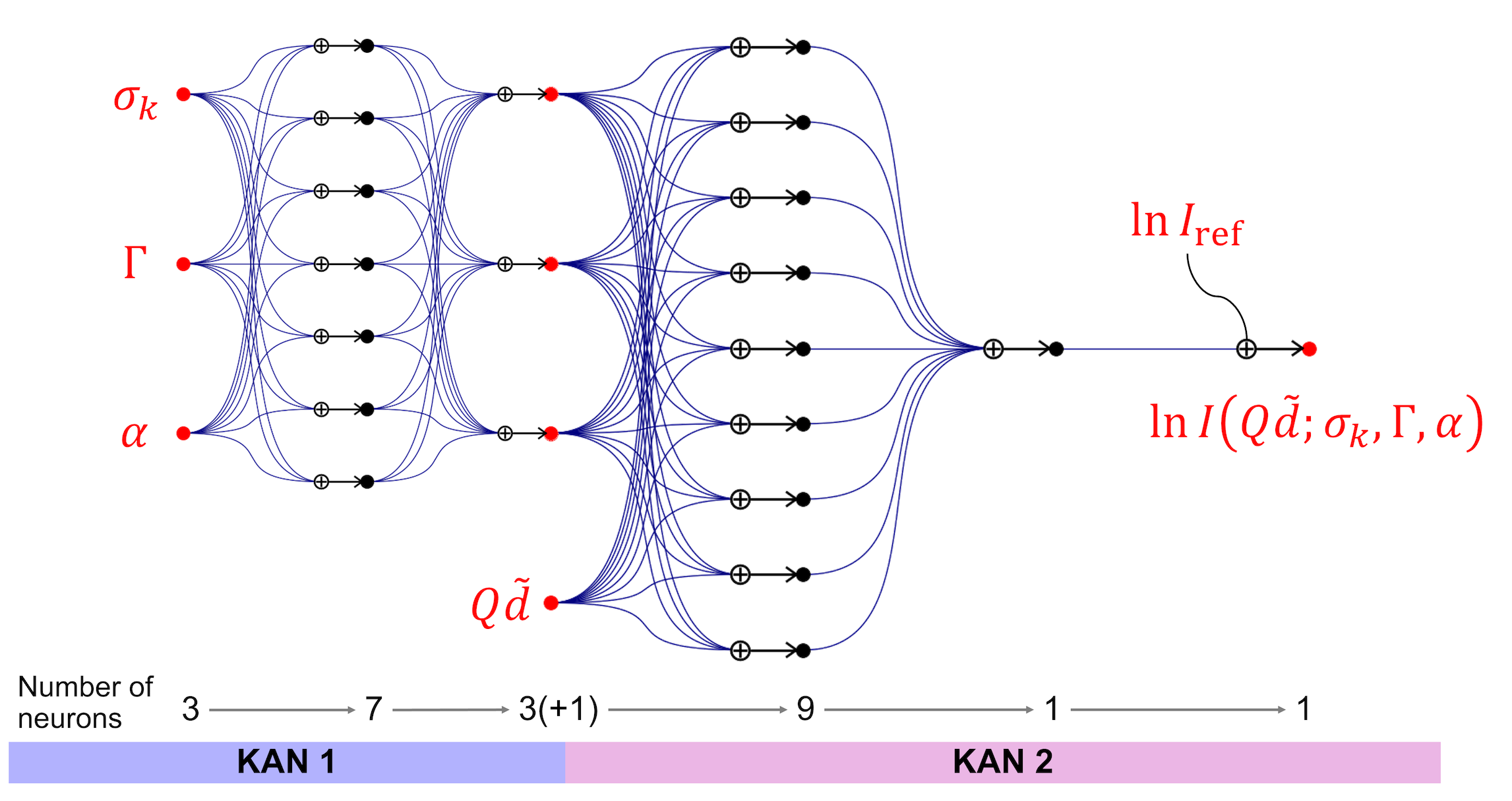}
  %\vspace{0 cm}
}  
\caption{KAN-based generative function for scattering intensity of defective lamellar phases. The network consists of two stages: KAN 1 (left) processes input parameters $\sigma_k$, $\Gamma$, $\alpha$, and $Q\tilde{d}$ through hidden layers with spline-based connections, each of the connecting curve are specified by the coefficients of spline functions. The output from KAN 1 feeds into KAN 2 (right), which further refines the features to predict the scattering intensity, $\ln{I(Q\tilde{d}; \sigma_k, \Gamma, \alpha)}$, after combining with the reference intensity, $\ln{I+\mathrm{ref}}$. The network architecture captures the nonlinear relationships between structural parameters and scattering profiles, enabling the inversion of experimental data by algorithm~\ref{algorithm_curvefit}.}
\label{fig:3}
  %\vspace{-0.5cm}
\end{figure}
%----------------------------------------------------

Fig.~\ref{fig:3} illustrates the KAN-based architecture designed to reconstruct the logarithmic scattering intensity, $\ln{I(Q\tilde{d}; \sigma_k, \Gamma, \alpha)}$, as a function of the wavevector $Q$ and structural parameters $\sigma_k$, $\Gamma$, and $\alpha$. This architecture employs the Kolmogorov-Arnold representation theorem to decompose the complex multivariate mapping into sums and compositions of univariate functions, implemented through a hierarchical neural network.

The input layer consists of the structural parameters $\sigma_k$, $\Gamma$, $\alpha$. These inputs are processed by KAN 1, which maps the input space to an intermediate feature space through layers containing 3, 7, and 3 neurons, respectively.
KAN 2 transforms the features from KAN 1 to predict the scattering intensity and incorporate the dependence on scaled wavevector $Q\tilde{d}$, where $\tilde{d}$ represents the average inter-planar spacing. This stage reduces the dimensionality progressively with layers containing 9, 1, and 1 neuron, ultimately outputting a scalar value, $\ln{I(Q\tilde{d}; \sigma_k, \Gamma, \alpha)}$. The final function linking the output \( \ln{I(Q\tilde{d}; \sigma_k, \Gamma, \alpha)} \) to its parent node serves as a flexible buffering, ensuring more numerically stable generation of the oscillating features in lamellar scattering functions. Each curve connecting two nodes correspond to one univariate function represented by spline, providing flexibility in modeling the interactions among structural parameters.

One of the feature of the spline basis within the KAN layers is their natural tendency to produce outputs that oscillate around zero~\cite{Spline}. To capitalize on this property, a standardization process using a suitably chosen baseline function significantly enhances the model's ability to identify and emphasize critical variations in scattering data.

In the training set~\cite{CNN2}, the scattering intensity \( \ln{I(Q\tilde{d})} \) evolves systematically with variations in \(\sigma_k\), \(\Gamma\), and \(\alpha\). While no simple analytical expressions exist for these variations, their trends suggest that preprocessing the training data through a linear algebra approach is both feasible and beneficial.

Providing the training set \(\{\mathbf{\lambda}_\mathrm{train}, \mathbf{I}_\mathrm{train}\}\), with shape of \(N_\mathrm{train}\times 3\) and \(N_\mathrm{train}\times N_Q\), the standardization process begins with a linear approximation for any \(1\times 3\) input parameter \(\lambda\):
\[
\ln{I_{\lambda}} = (\mathbf{\lambda}\mathbf{A} + \mathbf{B})^\top,
\]
where the matrices \(\mathbf{A}\) and \(\mathbf{B}\) are determined from the training set as follows:
\begin{equation}
\mathbf{A} = \left(\mathbf{\lambda}_\mathrm{train}^\top \mathbf{\lambda}_\mathrm{train}\right)^{-1} \mathbf{\lambda}_\mathrm{train}^\top \mathbf{I}_\mathrm{train},
\end{equation}
\begin{equation}
\mathbf{B} = \frac{1}{N_\mathrm{train}} \sum_{i=1}^{N_\mathrm{train}}[\mathbf{I}_{\mathrm{train}}-\mathbf{\lambda}_{\mathrm{train}}\mathbf{A}]_i.
\end{equation}
Here, \(\mathbf{A}\) is an \( 3 \times N_Q \) matrix, and \(\mathbf{B}\) is \( 1 \times N_Q \), where \( N_Q \) is the number of training set \( Q \)-values. For any input parameter vector \(\mathbf{\lambda}\), the corresponding intensity \( \ln{I_{\lambda}} \) of length \( N_Q \) can then be computed at the fixed \( Q \)-sampling points.

To accommodate arbitrary \( Q \)-values within the range of the training set, variations in the referential intensity, \( I_\mathrm{ref}(Q) \), are modeled using the following quadratic approximation:
\[
\ln{I_\mathrm{ref}(Q)} = \mathbf{C}^\top X_Q,
\]
where the \(3 \times 1\) coefficients \(\mathbf{C}\) are determined as:
\begin{equation}
\mathbf{C} = \left(X_\mathrm{train}^\top X_\mathrm{train}\right)^{-1} X_\mathrm{train}^\top \ln{I_{\lambda}}.
\end{equation}
The matrix \( X_\mathrm{train} \) is defined as:
\begin{equation}
X_\mathrm{train} = 
\begin{bmatrix}
Q_{\text{train},1}^2 & Q_{\text{train},1} & 1 \\
\vdots & \vdots & \vdots \\
Q_{\text{train},N_Q}^2 & Q_{\text{train},N_Q} & 1 \\
\end{bmatrix},
\end{equation}
and \( X_Q \) for any specific \( Q \)-value is given by:
\begin{equation}
X_Q = 
\begin{bmatrix}
Q^2 & Q & 1
\end{bmatrix}.
\end{equation}

The final preprocessed training data to be reproduced by KAN in Fig.~\ref{fig:3} is obtained by subtracting the reference logarithmic intensity:
\[
\ln{I(Q\tilde{d})} - \ln{I_\mathrm{ref}(Q)}.
\]

%--------------------------------------------------
\begin{figure}[htbp]
\centerline{
  \includegraphics[width=\linewidth]{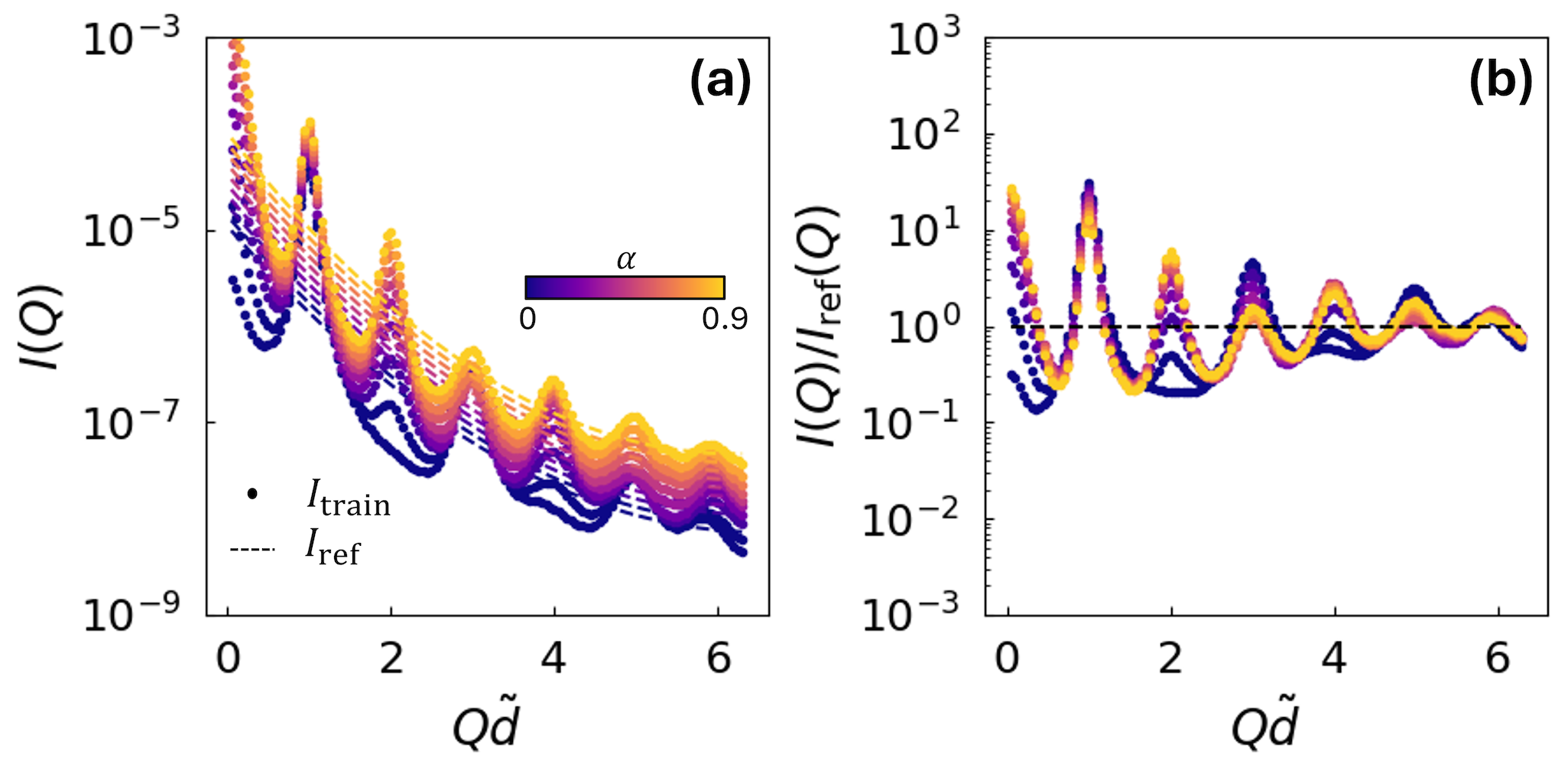}
}  
\caption{(a) Scattering intensity \( I(Q) \) as a function of the rescaled wave vector \( Q\tilde{d} \), where the training data \( I_{\mathrm{train}} \) (dots) and the reference intensity \( I_{\mathrm{ref}} \) (dashed lines) are shown for varying levels of structural distortion parameter \(\alpha\). The color gradient corresponds to increasing values of \(\alpha\). (b) Standardized intensity \( I(Q)/I_{\mathrm{ref}}(Q) \), where deviations from unity highlight residual differences caused by structural distortions. The standardization effectively normalizes the data, allowing the KAN model to capture and reconstruct variations in scattering intensity with high accuracy.}

\label{fig:4}
\end{figure}
%----------------------------------------------------

The implementation of this standardization process is illustrated in Fig.~\ref{fig:4}. Panel (a) shows the scattering intensity \( I(Q) \) as a function of the rescaled wave vector \( Q\tilde{d} \), where training data \( I_{\mathrm{train}} \) (dots) are compared with the reference intensity \( I_{\mathrm{ref}} \) (dashed lines) across increasing levels of the structural distortion parameter \(\alpha\). The gradual deviations from \( I_{\mathrm{ref}} \) highlight the impact of \(\alpha\) on the scattering profiles. Panel (b) displays the standardized intensity \( I(Q)/I_{\mathrm{ref}}(Q) \), where fluctuations around unity emphasize the residual differences due to structural distortions. This normalization ensures the KAN model robustly identifies and reconstructs variations in scattering intensities across the training set. The KAN model is then trained to capture the relations between \( I_{\mathrm{train}} \) and \(\{\sigma_k, \Gamma, \alpha\}\) in the training set~\cite{CNN2}. 

%----------------------------------------------------
\begin{figure}[htbp]
\centerline{
  \includegraphics[width=\linewidth]{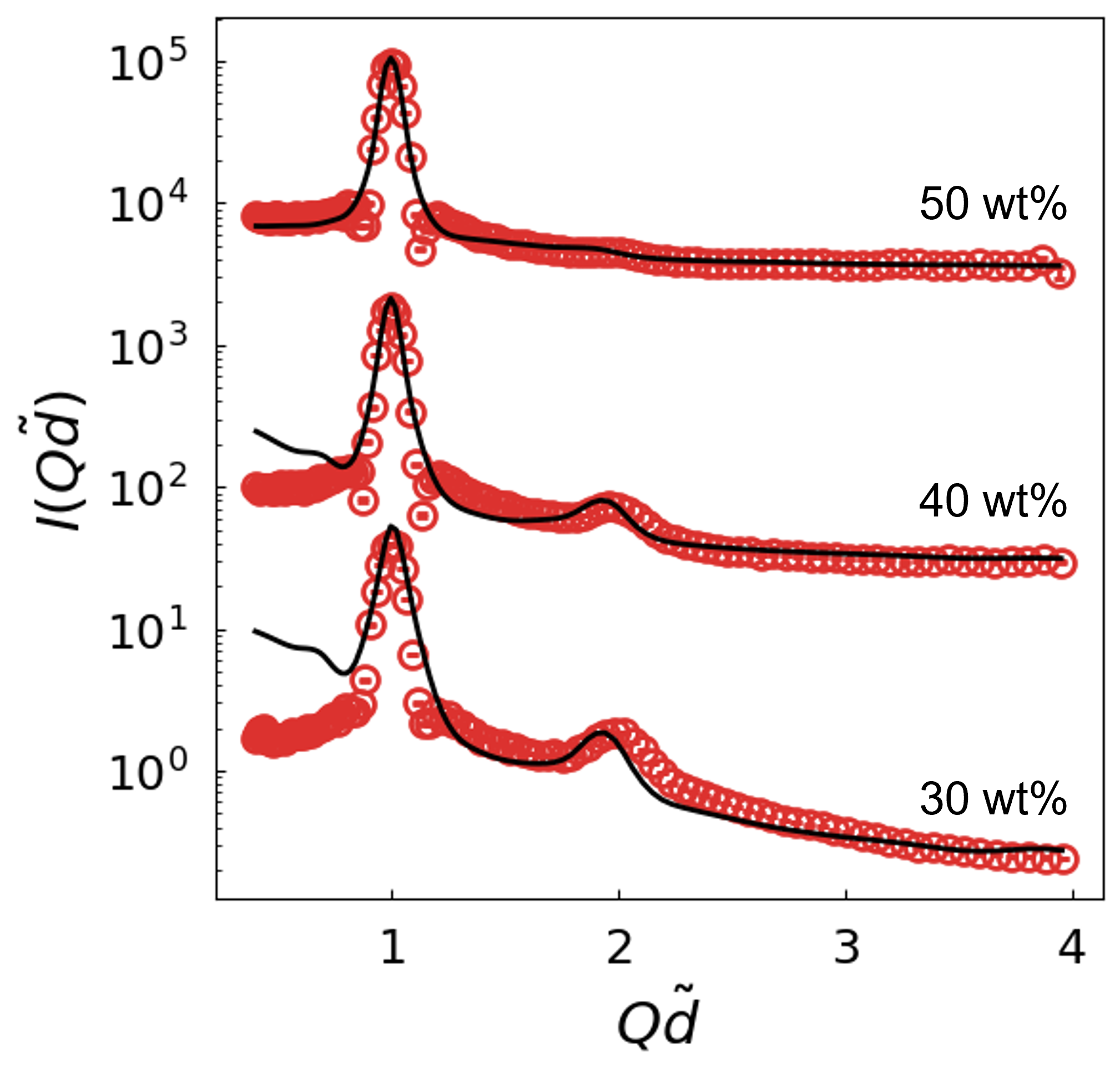}
  %\vspace{0 cm}
}  
\caption{SANS intensity, $I(Q\tilde{d})$, for aqueous solutions of sodium dioctyl sulfosuccinate (AOT) at 30\%, 40\%, and 50\% concentrations, presented in dimensionless units with $Q\tilde{d}$ as the key parameter. The symbols and the lines represent the experimental data and the KAN model, respectively. The average inter-plane distance, $\tilde{d}$, is estimated as $\frac{2\pi}{\tilde{Q}}$, where $\tilde{Q}$ is the position of the first correlation peak. As AOT concentration increases, the first correlation peak decreases in height and shifts to higher $Q$ values, with distinct changes in lamellar structural organization, particularly in the second correlation peak.}
\label{fig:5}
  %\vspace{-0.5cm}
\end{figure}
%----------------------------------------------------

We assess the feasibility of applying the KAN model to  algorithm~\ref{algorithm_curvefit} for SAS data analysis of defective lamellar systems. To this end, we choose of a well-established lyotropic system, sodium dioctyl sulfosuccinate (AOT) \cite{AOT}, as example. AOT molecules were dispersed in deuterium oxide (D$_2$O) to prepare aqueous solutions with concentrations of 30\%, 40\%, and 50\% by weight. SANS measurements were carried out on the D22 small-angle diffractometer at the Institut Laue-Langevin (ILL), employing two neutron wavelengths (6 \r{A} and 11.5 \r{A}) to cover a $Q$-range from 0.001 \r{A}$^{-1}$ to 0.5 \r{A}$^{-1}$, capturing coherent neutron scattering.

Fig.~\ref{fig:5} presents the SANS intensity, $I(Q\tilde{d})$, for aqueous solutions containing 30\%, 40\%, and 50\% AOT. The intensity is plotted in dimensionless $Q\tilde{d}$. As the AOT weight fraction increases, qualitative trends indicate a reduction in the height of the first correlation peak in $I(Q\tilde{d})$ for $Q\tilde{d}\sim 1$ and a shift in its position towards higher $Q\tilde{d}$ values. These changes suggest modifications in the structural organization of the lamellar phases, consistent with the AOT/water phase diagram~\cite{AOT}. Additionally, variations in the height and width of the second correlation peak in $I(Q\tilde{d})$ further reflect these conformational evolutions.

Quantitative analysis shows that experimental data (symbols) strongly agree with the KAN model (black curves) in the high-$Q$ regime, particularly for $Q\tilde{d} \gtrsim 1$, with minor deviations near the second correlation peak at $Q\tilde{d} \sim 2$. In contrast, significant discrepancies arise in the low-$Q$ regime ($Q\tilde{d} < 1$), where the experimental data flatten, while the KAN model predicts an upturn in $I(Q\tilde{d})$ that diminishes with increasing AOT concentration.

The origins of these discrepancies can be explored within the wave field representation \cite{CNN2} for distorted lamellar phases. Instrument resolution, addressed in prior work~\cite{desmear}, is unlikely the source. In the high \( Q \tilde{d} \) region, we hypothesize that minor discrepancies stem from the ansatz used to quantify the wave vector distribution, particularly regarding the formulation of \( P(k) \). An alternative approach, involving a zero mean but non-zero second and fourth moments~\cite{Choi}, contrasts with our method, which assumes a non-zero mean~\cite{Berk2}.

Additionally, a more general formulation for \( P(\theta) \), based on different Legendre polynomials~\cite{Arfken}, could allow greater flexibility in describing detailed angular dispersion. This approach, however, would introduce more parameters, requiring a larger training dataset and increasing computational cost, reflecting the "curse of dimensionality"~\cite{Bishop}.

The low \( Q \) disagreement may result from sample loading affecting grain orientation distributions in AOT solutions. As noted by Kekicheff et al.~\cite{Kekicheff1}, this process can alter the orientational distribution of grains in lamellar phases, influencing low-$Q$ scattering. In contrast, a sponge phase is unaffected by loading, explaining why discrepancies are most noticeable in 30\% AOT solutions, less so in 40\% AOT, and virtually absent in 50\% AOT solutions with sponge-like topology.

%--------------------------------------------------
\begin{figure}[htbp]
\centerline{
  \includegraphics[width=\linewidth]{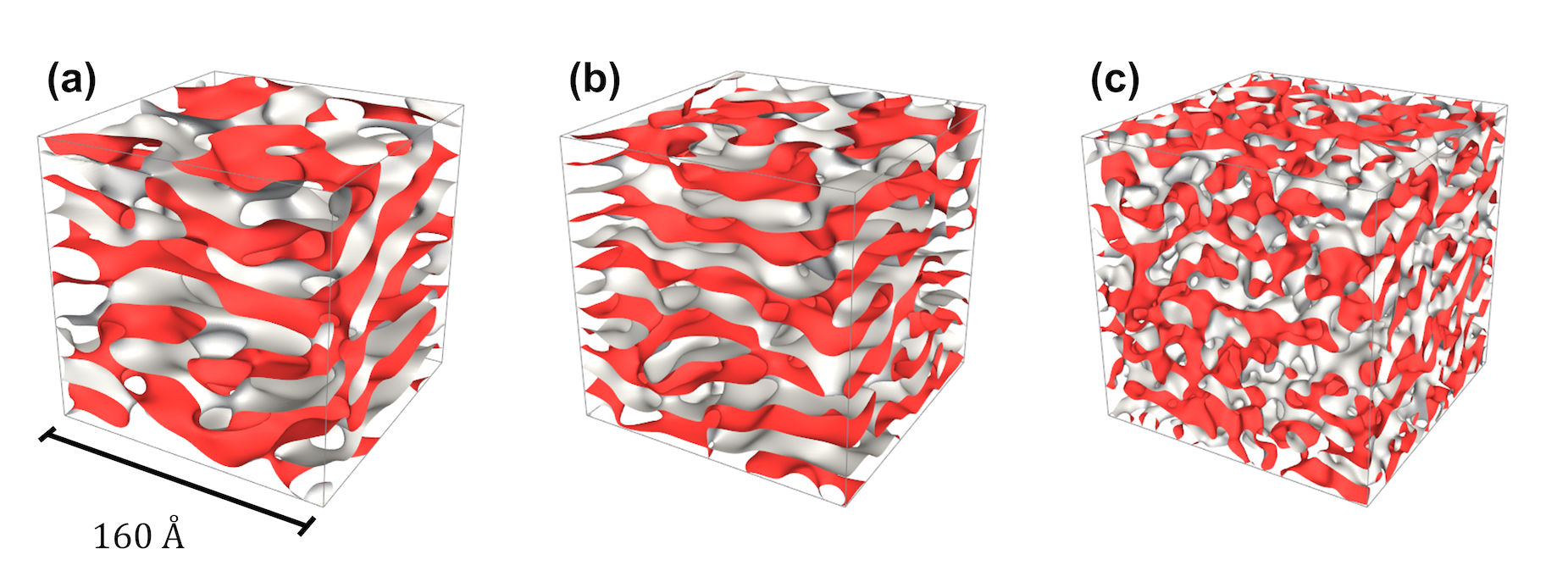}
  %\vspace{0 cm}
}  
\caption{Three-dimensional visualization of the structures derived from the extracted values of $\sigma_k$, $\Gamma$, and $\alpha$ for (a) 30\%, (b) 40\%, and (c) 50\% AOT/water solutions. The inter-planar spacing is on the order of tens of \r{A}. As the AOT weight fraction increases, pathways between adjacent plates emerge, enhancing the interlayer connectivity. This progression leads to the transformation of initially anisotropic, two-dimensional plate structures into a more isotropic phase at the mesoscopic scale.}
\label{fig:6}
  %\vspace{-0.5cm}
\end{figure}
%----------------------------------------------------

From the values of \( \sigma_k \), \( \Gamma \), and \( \alpha \), a three-dimensional wave field representation of lamellar structure (Fig.~\ref{fig:6}) is constructed. As AOT concentration increases, pathways between adjacent plates develop, transitioning from anisotropic plates to a more isotropic phase. Real-space renderings of these distorted lamellar phases, including inter-layer distances and in-plane correlation lengths, are presented in Fig.~\ref{fig:6}, with further details in Ref.~\cite{CNN2}.

\subsection{Charged Colloidal Suspensions}

The stability of charged colloidal suspensions is governed by the interplay between electrostatic screening and van der Waals forces, as described by the classical Derjaguin-Landau-Verwey-Overbeek (DLVO) theory \cite{DLVO}. The electrostatic interactions comprise a repulsive force from the overlap of ionic double layers and an attractive Hamaker interaction, typically on the order of $k_{\mathrm{B}}T$, where $k_{\mathrm{B}}$ is the Boltzmann constant and $T$ the temperature. Double-layer repulsion often dominates in low ionic strength systems, influencing particle arrangement via long-range interactions. According to scattering theory, interactions decaying faster than $1/r$ are short-ranged, while those decaying as $1/r$ or slower, such as electric double-layer forces, are long-ranged \cite{Isa}.

The screened Coulomb repulsion in charged colloids is commonly modeled by a Yukawa potential derived from the Debye-H\"uckel equation \cite{DLVO}:
\begin{equation}
\beta\; V_{R}(r) =
\begin{cases}
\infty, & \text{if } r < D,\\
A \dfrac{\exp\left[ -\kappa(r-D) \right]}{r}, & \text{otherwise},
\end{cases}
\label{eq:Yukawa}
\end{equation}
where $\beta = 1/k_{\mathrm{B}}T$, and the potential strength $A$ depends on the charge number $Z$, the electric charge $e$, solvent dielectric constant $\epsilon$, inverse Debye length $\kappa$, and colloid diameter $D$. The Yukawa potential effectively captures the electrostatic behavior of diverse charged colloidal systems, providing insights into their properties and interactions.

SAS is a vital technique for probing the effective interparticle potential $V_{R}(r)$, with Henderson's theorem linking this potential to static correlation functions such as the structure factor $S(Q)$ and the radial distribution function $g(r)$ \cite{Henderson, Gray, Egelstaff}. The inversion of $S(Q)$ to $V_{R}(r)$ is traditionally approached through integral equation (IE) theories, such as the Ornstein-Zernike (OZ) equation \cite{OZ, Hansen}:
%--------------------------------------
\begin{equation}
h(r_{12}) = c(r_{12}) + n_{p} \int c(r_{13}) h(r_{23}) \, d^{3}r_{3},
\label{eq:OZ}
\end{equation}
%--------------------------------------
where $h(r_{12}) \equiv g(r_{12}) - 1$, $c(r_{12})$ is the direct correlation function, and $n_{p}$ is the particle number density. Here, $S(Q)$ is the Fourier transform of $g(r)$. Solving Eqn.~\eqref{eq:OZ} requires a closure equation, but common closures—such as the mean spherical approximation (MSA) \cite{MSA}, rescaled MSA (RMSA) \cite{RMSA}, and penetrating-background RMSA (PB-RMSA) \cite{RMSAPB}—struggle with highly concentrated or strongly charged systems \cite{BeresfordSmith:1985, ILL, Klein:DAguanno:1996, MPBRMSA}. While the modified PB-RMSA (MPB-RMSA) \cite{MPBRMSA} improves accuracy, it faces convergence issues for $1/\kappa D < 0.1$ or $\phi > 0.3$ and becomes computationally inefficient at high volume fractions.

Other closures, such as the hypernetted chain (HNC) and Percus-Yevick (PY) approximations \cite{Schmitz}, systematically deviate: HNC underestimates $S(Q)$, while PY overestimates it \cite{Belloni:1991}. The Rogers-Young (RY) closure \cite{RY} performs better for moderately charged systems but remains computationally demanding and lacks robust convergence.\cite{Hus:2013} In strongly coupled systems with intense colloidal and counterion interactions, IE theories fail entirely,\cite{Anta:2002} highlighting their limitations for inverting $V_{R}(r)$ in complex, collective systems.\cite{Lee}

This report contends that achieving a perfect closure for solving the inverse scattering problem is unattainable. The pursuit of another analytically complex closure is unlikely to resolve this challenge. Instead, we propose a machine learning approach based on the KAN, which exploits the expressive power of scattering functions to establish a direct and efficient mapping between scattering data and potential parameters. As demonstrated in the inverse problem of defective lamellar phases, this method provides a promising regression framework for analyzing \( S(Q) \) in charged colloidal suspensions.

In the context of Eqn.~\eqref{eq:OZ}, the structure factor $S(Q)$ can be expressed as \cite{Schmitz, Hansen}:
\begin{equation}
S(Q) = \frac{1}{1 - \phi \tilde{c}(Q)},
\end{equation}
where $\tilde{c}(Q)$ is the Fourier transform of the direct correlation function $c(r)$ as defined in Eqn.~\eqref{eq:OZ}. 

Similarly to the case study of defective lamellar phases, it is necessary to identify a referential system to facilitate the KAN-based generative function. For this purpose, the hard sphere (HS) system is selected, as it represents an asymptotic limit for charged colloidal suspensions when \(1/\kappa D \rightarrow 0\) and \(A \rightarrow 0\) \cite{GPR1}. Using the PY closure \cite{Schmitz}, the structure factor $S(Q)$ of HS systems, derived from solving Eqn.~\eqref{eq:OZ}, takes the following form~\cite{HSPY}:
\begin{equation}
S(Q, R, \phi) = \frac{1}{1 + 24 \phi \frac{G_\mathrm{HS}(K)}{K}},
\end{equation}
where $K = 2QR$, and $G_\mathrm{HS}(K)$ is given by:
%---------------------------------------
\begin{equation}
\begin{split}
G_\mathrm{HS}(K) &= \alpha\frac{\sin K - K \cos K}{K^2} \\
&\quad+ \beta\frac{2K \sin K + (2 - K^2) \cos K - 2}{K^3} \\
&\quad+ \gamma\frac{-K^4 \cos K}{K^5} \\
&\quad+ \gamma\frac{4 \left[(3K^2 - 6) \cos K + (K^3 - 6K) \sin K + 6\right]}{K^5}.
\end{split}
\end{equation}
%------------------------------------
The following expressions define the parameters \(\alpha\), \(\beta\), and \(\gamma\):
%------------------------------------------
\begin{equation}
\alpha = \frac{(1 + 2\phi)^2}{(1 - \phi)^4},
\end{equation}

\begin{equation}
\beta = -6 \frac{(1 + \phi / 2)^2}{(1 - \phi)^4},
\end{equation}

\begin{equation}
\gamma = \frac{\phi (1 + 2\eta)^2}{(1 - \phi)^4}.
\end{equation}
%----------------------------------------
Based on this representation, we can train the KAN to capture the differences between the Yukawa potential $\tilde{c}(Q)$ and that of the referential HS system. The $S(Q)$ can then be expressed as:
\begin{equation}
S_\mathrm{KAN}(Q, R, \phi) = \frac{1}{1 + 24 \phi \frac{G_\mathrm{HS}(K) + \Delta G(K)}{K} + f_3(K)},
\label{eq:S_KAN}
\end{equation}
where \(\Delta G(K)\) represents the deviation from the hard sphere model, and \(f_3(K)\) is an additional correction term.

%--------------------------------------------------
\begin{figure}[htbp]
\centerline{
  \includegraphics[width=\linewidth]{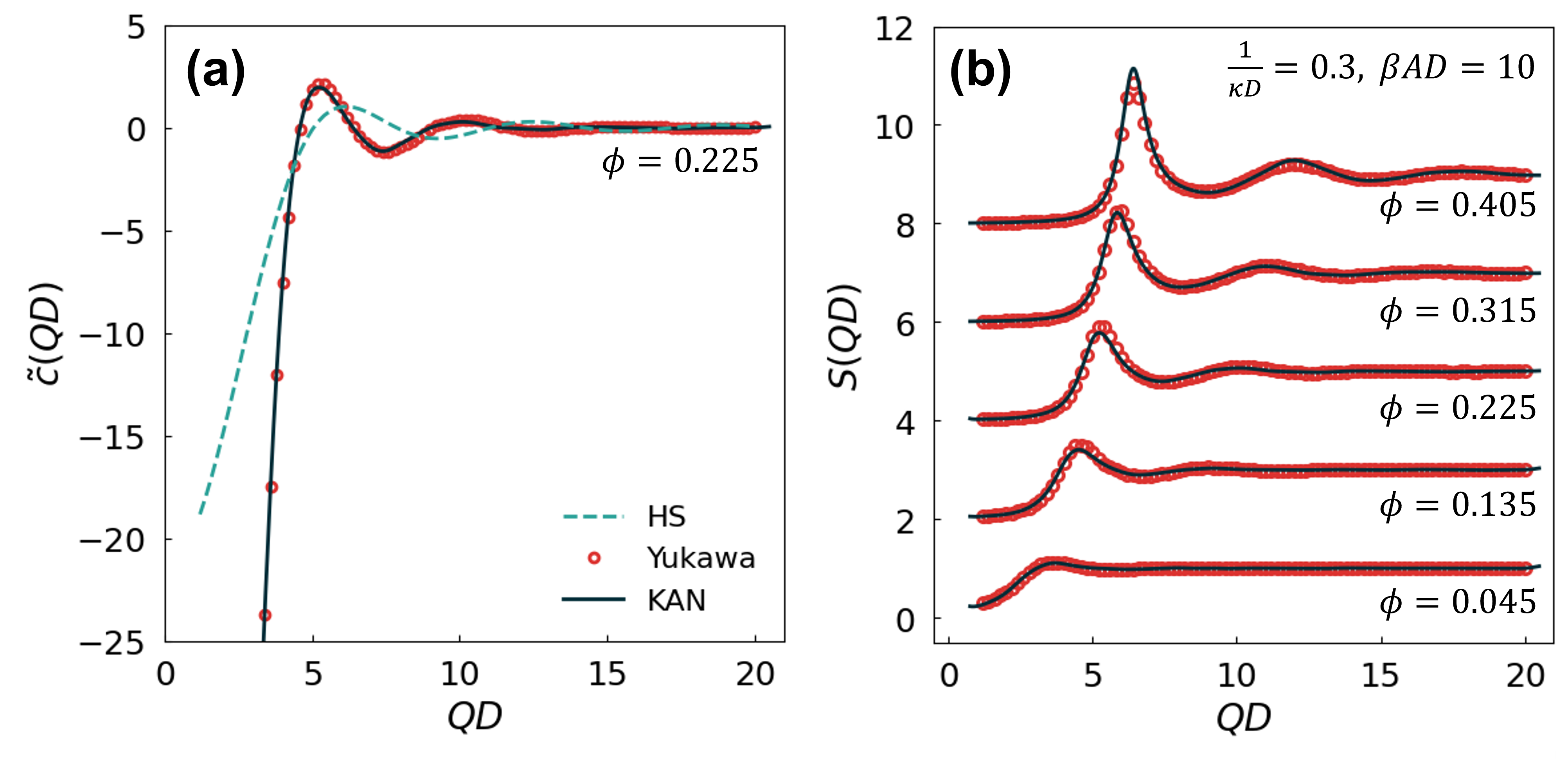}
  %\vspace{0 cm}
}  
\caption{Panel (a) compares the direct correlation function from the MD-generated training set (red circles) with the hard sphere (HS) system (cyan dashed curve), highlighting oscillations in the discrepancy \(\Delta G(K)\) at lower \(Q\). The direct correlation function is modeled as \(\tilde{c}_\mathrm{KAN}(K) = -24 \frac{G_\mathrm{HS}(K) + \Delta G(K)}{K} + \frac{f_3(K)}{\phi}\), and the KAN-generated result (black solid curve) shows excellent agreement with the MD data. In Fig.~\ref{fig:8} (b), the static structure factor \(S(Q)\) from MD simulations (red symbols) is compared with the KAN predictions (black solid lines) across a range of volume fractions \(\phi\), demonstrating quantitative agreement and confirming the KAN’s accuracy and efficiency in regression analysis for charged colloidal suspensions.
}
\label{fig:7}
  %\vspace{-0.5cm}
\end{figure}
%----------------------------------------------------

Fig.~\ref{fig:7} (a) illustrates the direct correlation function generated from the molecular dynamics (MD)-generated training set (red circles) in comparison with the results from a hard sphere (HS) system characterized by the same volume fraction \(\phi\) and particle radius \(R\) (cyan dashed curve). It is evident from observation that the discrepancy \(\Delta G(K)\) exhibits pronounced oscillations at lower wave vectors \(Q\). To describe this behavior, we model \(\Delta G(K)\) as:
\begin{equation}
\Delta G(K) = f_1(K)\sin(f_2(K))/K,
\end{equation}
where \(f_1(K)\) and \(f_2(K)\) represent functional dependencies on the wave vector. Based on this relationship, the direct correlation function is expressed as:
\begin{equation}
\tilde{c}_\mathrm{KAN}(K) = -24 \frac{G_\mathrm{HS}(K) + \Delta G(K)}{K} + \frac{f_3(K)}{\phi}.
\end{equation}

Here, \(G_\mathrm{HS}(K)\) corresponds to the direct correlation function of the hard sphere reference system, and \(f_3(K)\) introduces an additional correction term dependent on the volume fraction \(\phi\). 

%--------------------------------------------------
\begin{figure}[htbp]
\centerline{
  \includegraphics[width=\linewidth]{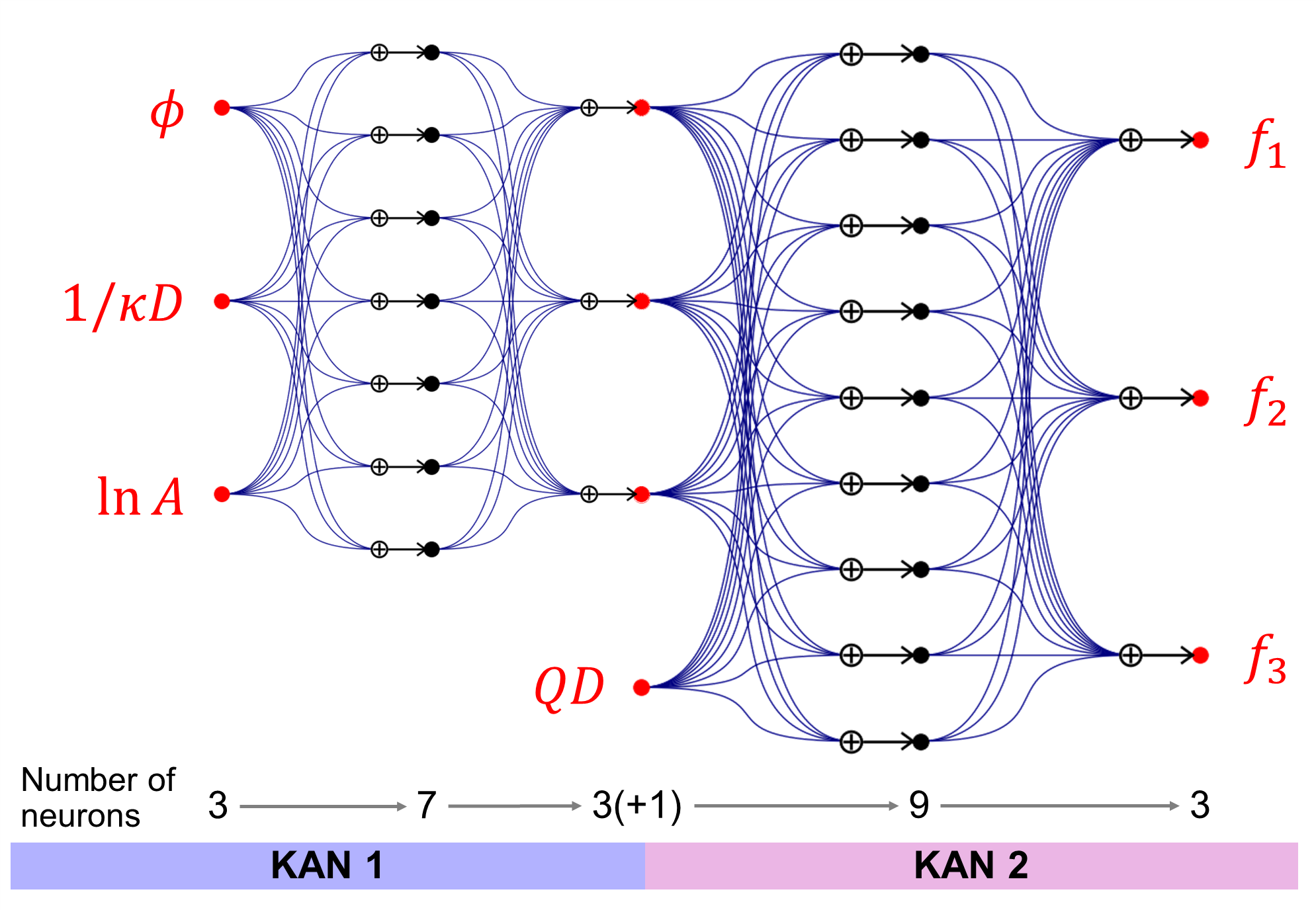}
  %\vspace{0 cm}
}  
\caption{Kolmogorov-Arnold Network (KAN) architecture for regression analysis of scattering data. The input parameters (\(\phi\), \(1/\kappa D\), \(\ln A\), \(QD\)) are mapped through two sub-networks (KAN 1 and KAN 2) to predict correction terms \(f_1\), \(f_2\), and \(f_3\). The number of neurons in each layer is indicated.}
\label{fig:8}
  %\vspace{-0.5cm}
\end{figure}
%----------------------------------------------------

The KAN architecture shown in Fig.~\ref{fig:8} illustrates a generative function for building \(S_\mathrm{KAN}\) in Eqn.~\ref{eq:S_KAN}. The input layer consists of four parameters: the volume fraction \(\phi\), the inverse screening length \(1/\kappa D\), the logarithmic amplitude \(\ln A\), and the dimensionless scaled wave vector \(QD\). These serve as key descriptors that encode the physical characteristics of the system. The architecture is split into two connected sub-networks, KAN 1 and KAN 2, with intermediate layers progressively increasing the number of neurons from 3 to 7, and subsequently transitioning to 3(+1), 9, and finally 3 neurons. This hierarchical design enables the network to capture intricate nonlinear relationships between the input parameters and the outputs \(f_1\), \(f_2\), and \(f_3\), which represent correction terms accounting for deviations from the HS referential model. By systematically training the KAN using the training set~\cite{GPR1} containing structure factor \(S(Q)\) and corresponding potential parameters, this architecture effectively maps the input physical parameters to the correction terms of inter-particle correlation, thereby enhancing the model's accuracy for describing the charged repulsive colloidal systems. 
The black solid curve in Fig.~\ref{fig:7}(a) represents the direct correlation function \(\tilde{c}_\mathrm{KAN}(K)\) generated by the KAN, showcasing excellent agreement with the MD-generated data (red circles). This result highlights the ability of the KAN to accurately reconstruct the direct correlation function.

In Fig.~\ref{fig:7}(b), we further compare the static structure factor \(S(Q)\) obtained from an independent test set (red symbols) with predictions from the optimized KAN (black solid lines). The comparisons are performed across a range of volume fractions \(\phi\), spanning from 0.045 to 0.405. Remarkably, the KAN-generated structure factors exhibit quantitative agreement with the MD results across the entire probed range, demonstrating the robustness and accuracy of the KAN framework.
These benchmarks demonstrate the KAN's efficacy in regression analysis for inverse problems in charged colloidal suspensions, offering a numerically accurate and efficient tool for analyzing soft matter systems.

Before proceeding, it is instructive to highlight the subtle distinction between the strategy employed in the current training process for establishing the KAN regression framework and our earlier machine learning approach for the potential inversion of colloidal suspensions \cite{GPR1, VAE, CNN1}. Unlike the previous approach, which directly links the evolution of potential parameters to the characteristic development of $S(Q)$, the present framework adopts a referential HS system. By addressing the difference between the $c(Q)$, which is more prominent in the low-$Q$ region, this method enables precise capture of the low-$Q$ coherent intensity. This property is crucial for determining the compressibility of the probed system and, consequently, the inter-colloidal effective interaction.

%--------------------------------------------------
\begin{figure}[htbp]
\centerline{
  \includegraphics[width=\linewidth]{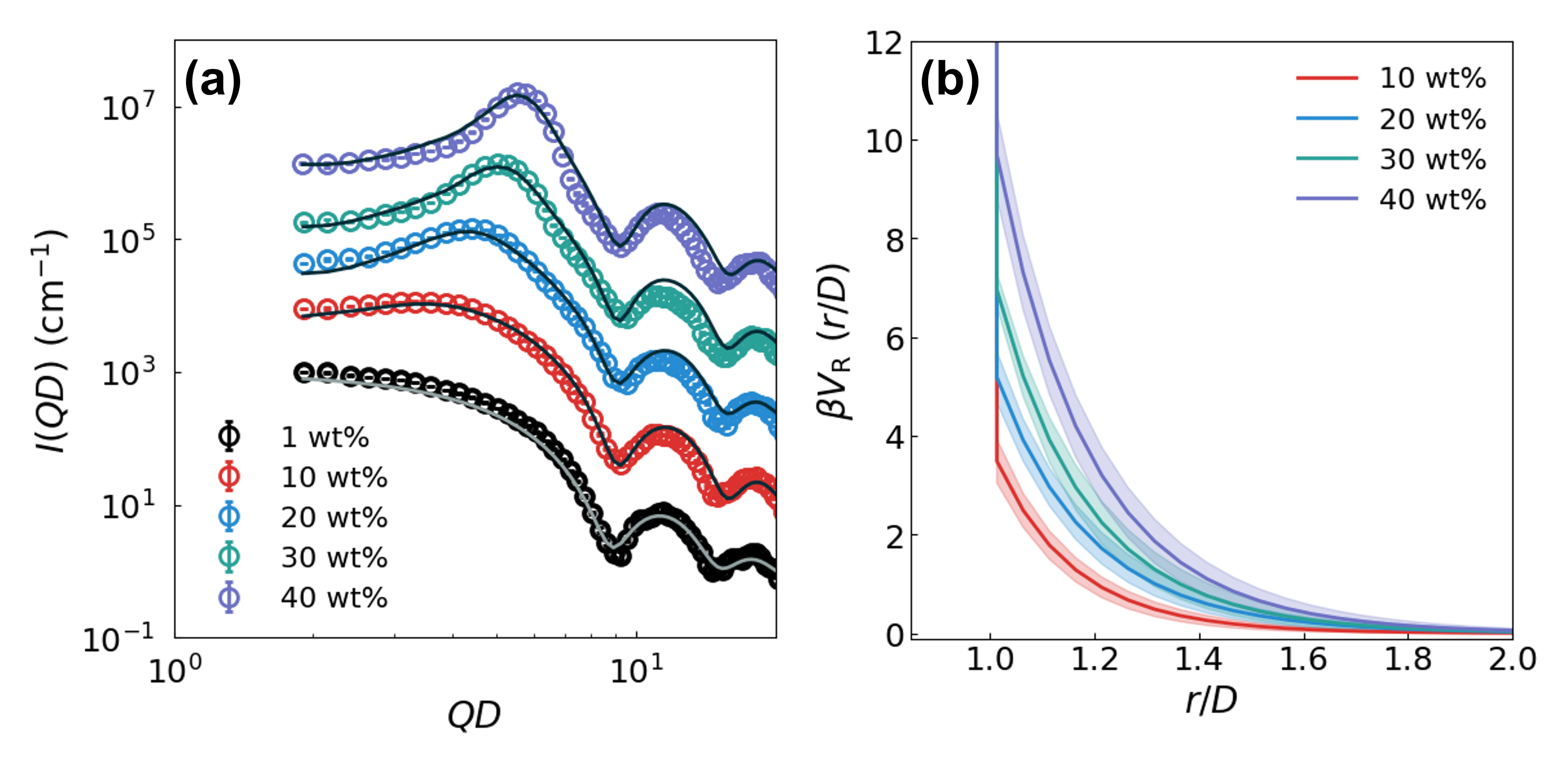}
  %\vspace{0 cm}
}  
\caption{(a) Scattering intensities, $I(QD)$, of silica colloidal dispersions at various weight fractions: 1 wt\% (black), 10 wt\% (red), 20 wt\% (blue), 30 wt\% (green), and 40 wt\% (purple). The particle diameter, $D$, was determined to be 115 nm based on the regression of the 1 wt\% data. To enhance clarity, the $I(QD)$ curves for each weight fraction have been offset vertically. (b) The inverse of the radial distribution function, $V_\mathrm{R}(\frac{r}{D})$, derived from the regression of the $I(QD)$ data. The observed fluctuations in $V_\mathrm{R}(\frac{r}{D})$ reflect the experimental uncertainty for each weight fraction.
}
\label{fig:9}
  %\vspace{-0.5cm}
\end{figure}
%----------------------------------------------------

Fig.~\ref{fig:9}(a) shows the SANS data (symbols) and the corresponding KAN predictions in the dimensionless representation of $QD$. The particle diameter, $D$, was determined to be 115 nm with 6\% polydispersity from 1 wt\% dilute dispersions (black symbols). At this polydispersity, the decoupling approximation can be used to calculate the scattering intensity as 
%-------------------------
\begin{equation}
\label{eq:decoupling}
    I(Q) = \left\langle P(Q)\right\rangle\left\langle S(Q)\right\rangle,
\end{equation}
%----------------------------
where $\left\langle P(Q)\right\rangle$ and $\left\langle S(Q)\right\rangle$ are the form and structure factors, respectively. The KAN-generated $S(Q)$ was used for curve fitting. The optimized $I(QD)$ obtained from algorithm~\ref{algorithm_curvefit} align well with the SANS data. Despite deviations near the structural peak increase at higher volume fractions due to the potential multiple scattering effect, the form factor oscillation shows that the approximation in Eqn.~\ref{eq:decoupling} remains valid for spherical particles with low polydispersity ($\sim6\%$).

% Volume fractions for samples from 10 to 40 wt\% show a linear increase with weight percentage. Using these to calculate the silica particle density reveals slight discrepancies with the measured value of 1.68 $g/\mathrm{cm}^3$.

Fig.~\ref{fig:9}(b) presents the inverted $V_\mathrm{R}(\frac{r}{D})$ curves, which show little variation across dispersions, with fluctuations within reasonable uncertainties. The repulsive strength is seen to increase upon increasing the silica weight fraction.

\section{Conclusions}
\label{sec:4}

This work introduces a machine learning framework based on the KAN to address the inverse problem in small-angle scattering of soft materials. By leveraging the KAN's ability to generate continuous functions, we successfully analyzed two representative soft matter systems: defective lamellar phases and charged colloidal suspensions. The KAN framework overcomes key limitations of traditional models, including the reliance on fixed $Q$ grids and the introduction of interpolation artifacts, while providing a robust and efficient approach to map structural parameters to scattering intensities directly for curve fitting.

For defective lamellar phases, the KAN-based model effectively captured the nonlinear relationships between structural parameters ($\sigma_k$, $\Gamma$, $\alpha$) and scattering intensity $I(Q)$. This capability enabled the resolution of structural distortions and the quantification of topological defect distributions. In the case of charged colloidal suspensions, the KAN established a mapping between structure factors $S(Q)$ and effective interaction potentials, achieving excellent agreement with both molecular dynamics simulations and experimental data. These findings underscore the versatility and accuracy of the KAN framework in reconstructing complex structural and interaction profiles in soft matter systems.

The application of the KAN framework to experimental data highlights its potential to extract meaningful structural insights while preserving the integrity of the data. Our results demonstrate that the KAN approach outperforms traditional analytical methods in both accuracy and adaptability, offering a robust alternative for analyzing scattering systems with variable $Q$-point distributions.

Looking ahead, the KAN framework presents exciting opportunities for advancing scattering-based analysis. Its flexibility to integrate additional physical constraints and extend to dynamic scattering techniques, such as neutron spin echo and x-ray photon correlation spectroscopy, offers a powerful avenue for probing non-equilibrium phenomena and time-resolved processes in soft materials. By establishing the KAN as a transformative tool for structural inversion, this work lays the foundation for broader applications in the characterization of materials with complex structures.

\begin{acknowledgments}
This research was performed at the Spallation Neutron Source and the Center for Nanophase Materials Sciences, which are DOE Office of Science User Facilities operated by Oak Ridge National Laboratory. This research was sponsored by the Laboratory Directed Research and Development Program of Oak Ridge National Laboratory, managed by UT-Battelle, LLC, for the U. S. Department of Energy and the U.S. Department of Energy Office of Science, Office of Basic Energy Sciences, Data, Artificial Intelligence and Machine Learning at DOE Scientific User Facilities Program under Award Number 34532. Monte Carlo simulations and computations used resources of the Oak Ridge Leadership Computing Facility, which is supported by the DOE Office of Science under Contract DE-AC05-00OR22725. Y.S. was supported by the U.S. DOE, Office of Science, Basic Energy Sciences, Materials Sciences and Engineering Division.
\end{acknowledgments}

\section*{Data Availability Statement}
The data that support the findings of this study are available from the corresponding author upon reasonable request.

\nocite{*}
\bibliography{ref}% Produces the bibliography via BibTeX.

\end{document}